\documentclass[12pt,letterpaper]{JHEP3}
\usepackage{amsmath,amssymb,amsfonts,xspace}

\newcommand{\de}{\mathrm{d}}

\newcommand{\I}{\mathrm{i}}

\newcommand{\cF}{\mathcal{F}}

\newcommand{\cM}{\mathcal{M}}

\newcommand{\cN}{\mathcal{N}}

\newcommand{\cO}{\mathcal{O}}

\newcommand{\cA}{\mathcal{A}}

\newcommand{\kahler}{{K\"ahler}\xspace}
\newcommand{\hk}{{hyperk\"ahler}\xspace}

\newcommand{\IR}{\mathbb{R}}
\newcommand{\IC}{\mathbb{C}}
\newcommand{\IZ}{\mathbb{Z}}

\newcommand{\Tr}{\mbox{Tr}}

\newcommand{\tzeta}{\tilde\zeta}

\newcommand{\Scl}{S_{\rm Cl}}

\def\bea{\begin{eqnarray}}
\def\eea{\end{eqnarray}}
\def\be{\begin{equation}}
\def\ee{\end{equation}}
\def\ba{\begin{align}}
\def\ea{\end{align}}
\def\bse{\begin{subequations}}
\def\ese{\end{subequations}}
\def\Im{\,{\rm Im}\,}
\def\Re{\,{\rm Re}\,}

\title{Large D-instanton effects in string theory}

\preprint{ITP-UU-09/15, SPIN-09/15}

\author{Boris Pioline$^{1}$, Stefan Vandoren$^2$
\\
$^1$ {\it Laboratoire de Physique Th\'eorique et Hautes
Energies, CNRS UMR 7589, \\
Universit\'e Pierre et Marie Curie,
4 place Jussieu, 75252 Paris cedex 05, France} \\

$^2$ {\it   Institute for Theoretical Physics and
           Spinoza Institute,
           Utrecht University,
           Leuvenlaan 4,
           3508 TD Utrecht,
           The Netherlands
           }

\vspace*{2mm} {\tt e-mail: \email{pioline@lpthe.jussieu.fr}, 
\email{S.J.G.Vandoren@uu.nl}} \vspace*{-3mm}

}

\abstract{By reduction along the time direction, black holes in 4 dimensions yield 
instantons in 3 dimensions. Each of these instantons  contributes  individually at order $\exp(-|Q|/g_s)$
to certain protected couplings in the three-dimensional effective action, but the number of 
distinct instantons is expected to be equal (or comparable) to the number of black hole 
micro-states, i.e. of order $\exp(Q^2)$. The same phenomenon also occurs 
for certain protected couplings 
in four dimensions, such as the hypermultiplet metric in type II string theories 
compactified on a Calabi-Yau threefold. In either case, the D-instanton series is 
therefore asymptotic, much like the perturbative expansion in any quantum field theory. By  using 
a Borel-type resummation method, adapted to the Gaussian growth of the D-instanton series, 
we find that the total D-instanton sum has an inherent
ambiguity of order $\exp(-1/g_s^2)$. We further suggest that this ambiguity can be 
lifted by including Kaluza-Klein monopole or NS5-brane instantons.
}

\begin{document}

The large order behavior of perturbation theory is a telltale hint on the nature of 
non-perturbative effects in quantum mechanics and quantum field theory \cite{Dyson:1952tj,Brezin:1976vw,Brezin:1976wa}. This also holds for string theory, and indeed, an estimate of the growth of 
string perturbation theory \cite{Gross:1988ib} led to the prediction of the existence 
of D-brane instantons \cite{Shenker:1990uf}
long before their actual construction \cite{Green:1994sx,Polchinski:1994fq,Polchinski:1995mt}.
D-instantons contribute to scattering amplitudes $\cA$ 
in string theory on $\IR^{1,d-1}\times Y$ schematically as 
\be 
\label{singleestim}
\cA_{\rm inst}(g_s, t^a, \theta_I) = \sum_{Q^I\in L}
\mu(Q^I, g_s, t^a)\,\exp\left( - \frac{1}{g_s} \Scl(Q,t^a)+ 2\pi \I \, \theta_I Q^I \right)\ ,
\ee
where  $Q^I$ are the Ramond-Ramond charges in $d$ dimensions, valued in some
rank~$n$ lattice $L$,  $\theta_I$  are the 
Ramond-Ramond axions, $t^a$ are the Neveu-Schwarz moduli, 
$\Scl(Q,t^a)$ is the classical action of the Euclideanized D-brane after extracting
one power of the string coupling $g_s$, and $\mu$ is a function
of $(Q,g_s,t^a)$ which behaves as a certain power of the 
string coupling constant as $g_s\to 0$ 
keeping the charges $Q^I$ and moduli $t^a$ fixed: 
\be
\label{mual}
\mu(Q,g_s,t^a)=g_s^\alpha \, \mu_0(Q,t^a) \left(1+\cO(g_s) \right)\ .
\ee 
Such instanton effects are typically 
negligible compared to perturbative corrections  at small coupling $g_s$, but may become 
dominant for certain processes where perturbative contributions are forbidden due to non-renormalization theorems.
In this note, we focus on ``BPS saturated'' couplings in the  effective action 
of superstring theory, which receive perturbative corrections only up to a certain genus,  
and non-perturbative corrections from  BPS instantons only, i.e. 
instantons (or multi-instantons) preserving a certain fraction of supersymmetry
(see e.g. \cite{Kiritsis:1999ss} for a review).

Our interest in this note is in the dependence of the ``instanton measure" $\mu_0(Q,t^a)$
on the charges $Q^I$, and in the convergence properties 
of the D-instanton series \eqref{singleestim}.  Since the classical
action $\Scl(Q,t^a)$ typically scales linearly with $Q^I$, any faster-than-linear growth
of $\log \mu$ as a function of the charges would imply that the series \eqref{singleestim}
would have zero radius of convergence, and should be treated as an asymptotic series. 

In ordinary quantum field
theory, $\mu_0(Q,t^a)$ can be calculated from the integration measure 
on the  instanton moduli space and  
the one-loop fluctuation determinants around the instanton
background\footnote{The exponent $\alpha$ in \eqref{mual} 
depends on the normalization of the external vertices, but usually not on $Q^I,t^a$;
its precise value is irrelevant for our purposes.}. For BPS instantons in supersymmetric field theories, the bosonic and fermionic fluctuation
determinants usually cancel, leaving only the integral of some characteristic class on
the instanton moduli space.  

In string theory, we do not know how to compute
$\mu_0(Q,t^a)$ from first principles. In certain cases however,
we may relate it to the indexed degeneracy of BPS solitons
as follows  \cite{Alexandrov:2008gh}.  Suppose that the compact manifold 
$Y$ is a product $X\times S^1$, and 
that the D-instanton in $\IR^d$ is obtained by wrapping a Euclideanized
D0-brane in $\IR^{d+1}$ along the Euclidean time circle $S^1$ of
radius $R$ (in particular, the D0-brane must have mass $M=\Scl(Q,t^a)/(2\pi R g_s)$
and electric and magnetic charges $Q^I$, so as to reproduce the instanton action \eqref{singleestim}; 
the D0-brane may itself
be obtained by wrapping D$p$-branes on some non-trivial $p$-cycle in $X$).
The instanton measure is given, up to a model-dependent normalization 
factor\footnote{We shall fix this proportionality factor in a specific example at the end of this note, 
when we discuss instantons in Calabi-Yau string compactifications.}, 
by\footnote{When the quantum mechanics has extended supersymmetry, one must include additional current insertions corresponding to the 
fermion bilinears appearing in the vertex $\cA$.} $\Tr[(-1)^{F} e^{-2\pi R H}]$ in the D0-brane
quantum mechanics with Hamiltonian $H$ \cite{Green:1997tv}.
Its  large radius limit $R\to \infty$ defines the Witten index $\Omega(Q)$, 
i.e. the indexed degeneracy of the D0-brane bound state in $\IR^{1,d}$. The latter 
is independent of both $g_s$ and $t^a$ by the attractor phenomenon
(though it may jump across lines of marginal stability). Moreover, 
when the spectrum is discrete, the trace is independent of $R$, and therefore
$\mu_0(Q,t^a)=\Omega(Q)$. If on the contrary the D-instanton in $\IR^d$ originates from
a D-instanton in $\IR^d \times S^1$ smeared along $S^1$, T-duality along this circle
maps it back to a Euclidean D0-brane wrapping $S^1$ of radius $l_s^2/R$, which 
reduces to a soliton in $\IR^{1,d}$ in the limit $R\to 0$. In this case again, $\mu(Q,t^a)$
becomes equal to the indexed degeneracy of the T-dual D0-brane. 
Thus, in either of these two cases, we have  \cite{Alexandrov:2008gh}
\be
\label{muom}
\mu_0(Q,t^a) \sim \Omega(Q) \ .
\ee
This relation may fail in cases where the D0-brane spectrum has a 
continuous part \cite{Sethi:1997pa,Yi:1997eg}.
This is for example the case of half BPS D-instantons in type IIB string theory   
in 9 dimensions, where the ``bulk'' contribution to the index precisely accounts
for the discrepancy between the two sides of \eqref{muom} \cite{Green:1998yf}. 
More generally, this is the case when the charge vector $Q^I$ is non primitive.
Similarly, 
we may expect that 
\eqref{muom} breaks down at a wall of marginal stability; on either side of the 
wall however, we expect that \eqref{muom} holds, as the same
jump should affect the index and the instanton measure \cite{Gaiotto:2008cd}.
For our present
purposes we shall only require that the two sides of \eqref{muom} have the same asymptotic growth.

It should also be noted that when $d=3$, there are additional instantons in 
$\IR^{3}\times S^1\times X$ which are of neither types above: Euclidean NS5-branes 
wrapped on $X$, and gravitational instantons  asymptotic to $\IR^{3}\times S^1$, also
known as Kaluza-Klein monopoles or KK5-branes. The action of these instantons scales as 
$\tau_2^2 V$ and $R^2 \tau_2^2 V$, where $V$ is the volume of $X$ in string units
and $1/\tau_2 \propto g_s$ is the ten-dimensional string coupling. Their
contributions are therefore exponentially
suppressed  compared to individual D-instanton contributions at weak coupling.
We shall return to these effects momentarily.

Granting \eqref{muom}, it is now straightforward 
to estimate the prefactor $\mu_0(Q,t^a)$ at large charge $Q$:
under the standard assumption that the index $\Omega(Q)$ is equal or comparable to the exact 
degeneracy at strong gravitational coupling, we can use 
the black hole representation of the D-brane configuration 
to conclude that 
\be
\label{muS}
\mu_0(Q,t^a) \sim\exp[S_{BH}(Q)] \ ,
\ee
where $S_{BH}(Q)$ is the Bekenstein-Hawking entropy. Our interest will be in 
situations where the gravitational solution is a single-centered 4D BPS black hole with 
a large horizon, tensored with a compact manifold $X$ (which may be a Calabi-Yau threefold,
$K3\times T^2$ or $T^6$). For the horizon
to be large in 4D Planck units, the 4D black hole must preserve no more than
4 supercharges. The coupling $\cA$ under study 
should therefore correspond to  a two-derivative coupling in 3D vacua with 8 
supercharges (e.g. the vector multiplet quaternionic metric in type II on $X\times S^1$), 
or a six-derivative coupling in a 3D vacua with 16 supercharges, or a fourteen-derivative 
coupling in 3D vacua with 32 supercharges. 
Even with this amount of supersymmetry, 
the existence of a single centered BPS solution typically requires some conditions on 
the total charge, e.g. $I_4(Q)>0$ in cases with 16 or 32 supercharges, where $I_4$
is the quartic polynomial such that $S_{BH}(Q)=\pi \sqrt{I_4(Q)}$. In the opposite case
($I_4(Q)<0$), there is usually no BPS black hole solution (although there may exist 
a non-BPS solution, not contributing to the index), and we set $S_{BH}(Q)=0$.
Combining \eqref{muom} and \eqref{muS}, we conclude that 
\be
\label{mua}
\mu_0(Q,t^a) = a(Q)\, \exp[S_{BH}(Q)] \ ,
\ee
where $a(Q)$ grows at most like a power of $Q$ at large $Q$.

In any of the cases above, the Bekenstein-Hawking entropy of a single-centered BPS 
black hole solution is homogeneous of degree 2 in the electric and magnetic charges 
$Q^I$, and therefore the instanton sum \eqref{singleestim} 
has zero radius of convergence~\footnote{This is in contrast to the usual Hagedorn divergence
in perturbative string theory, which leads to a pole in the partition
function.}.  This does 
not mean that it is useless, but rather that it must be regarded as an asymptotic expansion.
This is analogous  to the usual situation in 
quantum field theory , where the perturbative 
expansion, of the form\footnote{The case of perturbative string theory, corresponding to 
asymptotic series of the form $\cA(g_s)=\sum_n (2n)! \, a_n g_s^n$ \cite{Gross:1988ib,Shenker:1990uf},
can be treated in the same way, upon replacing $g \to \sqrt{g_s}$.}
$\cA(g)=\sum_{n\geq 0} n! \, a_n \,g^{2n}$ where $a_n$ is
bounded by some power $n^b$ of the loop order, is assumed to be the
asymptotic expansion of some non-perturbatively defined
function describing the exact answer for the amplitude ${\cal A}$  (see e.g. \cite{Beneke:1998ui} for a 
review). The truncated series 
${\cal A}_N(g)=\sum_{0\leq n\leq N-1} n! \, a_n \,g^{2n}$ should then approximate the 
exact result ${\cal A}(g)$ up to an error $\varepsilon$ which can be estimated to be of the order 
of the largest term in the sum, $\varepsilon=N! \, N^b \, |g|^{2N}$. This error is minimized upon choosing $N\sim 1/g^2$ for $g$ small and $N$ large.
At that optimum value, $\varepsilon \sim e^{-1/g^2}$, the inherent
ambiguity of the perturbative series.

Borel resummation consists in
representing $n! =\frac{1}{g^2} 
\int_0^\infty \de t\, (t/g^2)^{n} e^{-t/g^2}$ and exchanging the $\int$ and $\sum$ signs.
If the  Borel transform ${\cal B}(t)\equiv \sum_{n\geq 0} \, a_n \,t^{n}$ is well
defined and regular everywhere on the positive real axis, 
the series $\cA(g)$ is said to be ``Borel summable", 
and its Laplace transform $\frac{1}{g^2} \int_0^\infty e^{-t/g^2}{\cal B}(t)$ 
produces a function $\tilde \cA(g)$ with the same asymptotic expansion as ${\cal A}(g)$
in the sector $\Re(g^2)>0$.  
However, this procedure may be ambiguous due to singularities of 
${\cal B}(t)$ at particular points or branch cuts in the Borel $t$ plane, 
typically along the real $t$-axis. To define the Laplace transform, one must
choose a contour that avoids the singularities, but this choice of
contour is not unique. Different contours lead to answers that differ by terms
of order $\cO(e^{-1/g^2})$, and a full non-perturbative
definition of the quantum field theory is expected to fix these ambiguities, 
by relating them to computable instanton effects.

We can now apply the same line of reasoning 
to the divergent D-instanton series \eqref{singleestim}, where now the r\^ole
of $g$ is played by $e^{-1/g_s}$ and the growth of the Taylor coefficients is Gaussian rather than 
factorial. Our first task is to determine the optimal value of the cut-off on $Q^I$ such that the error
is minimized. Substituting \eqref{mua} into \eqref{singleestim} and dropping terms that scale
like powers of $Q$, we find that this is achieved when 
\be
\label{FullS}
\Sigma(Q,g_s,t^a) \equiv - S_{BH}(Q)  + \frac{1}{g_s} \Scl(Q,t^a)- 2\pi\I \, \theta_I Q^I\ ,
\ee
is maximized as a function of $Q$. Since $\Scl(Q,t^a)$ scales linearly with $Q$, the optimum
value of $Q$ is therefore of order $1/g_s$ at small $g_s$, making the ambiguity of the asymptotic series 
of order $\exp( -\kappa /g_s^2)$. 

To compute the coefficient $\kappa$, which will turn out to be positive, 
we need to specify the form of the instanton action $\Scl(Q,t^a)$.
For definiteness, we restrict to the case of 3D backgrounds with 8 supercharges, e.g.
type II string theory on $Y=X\times S^1$ where $X$ is a Calabi-Yau threefold.
In conventions where  $g_s$ is related
to the 10 dimensional string coupling $1/\tau_2$ via $1/g_s^2=8 R^2 V \tau_2^2$, 
the classical action of the 3D instanton,
or the BPS mass of the 4D black hole, is proportional to the 
modulus of the central charge of the $\cN=2$ Poincar\'e superalgebra, 
\be
\Scl(Q^I,t_a) = 2\pi\, |Z(Q)|\ ,\qquad
Z(Q)\equiv e^{K/2}\, Q^I \cF_I\ ,
\ee
where  $\cF_I=(X^\Lambda, F_\Lambda)$ is the holomorphic symplectic section
of $\cN=2$ supergravity and $K=- \log(\I  \cF_I \bar \cF^I)$ is the \kahler 
potential, related to the volume of $X$ in string units via 
$V=\frac18 e^{-K}$. Here and below, we use a notation in which indices are lowered using the symplectic form on $L\otimes \IC$, e.g. $\I \cF_I \bar \cF^I =\I(
X^\Lambda \bar F_\Lambda - F_\Lambda {\bar X}^\Lambda)$ . 

Moreover, the Bekenstein-Hawking entropy can be computed by
solving the ``attractor equations'' \cite{Ferrara:1995ih,Strominger:1996kf}
(see e.g. \cite{Pioline:2006ni} for a review), 
\be
\label{att}
\Re(\cF^I)= Q^I \quad \Rightarrow  \quad
S_{BH}(Q) = \frac{\I\pi}{4} \cF_I \bar \cF^I \geq 0 \ .
\ee
To linearize the optimization problem over $Q^I$, 
we introduce a ``twistor coordinate'' $z$  \cite{Neitzke:2007ke} and replace \eqref{FullS} by
\be
\label{Sigz}
\Sigma(Q,g_s,t^a,z) \equiv - S_{BH}(Q)  + \frac{\I\pi\, e^{K/2}}{g_s} 
\left( \cF_I\, z^{-1} - \bar \cF_I z \right) Q^I  - 2 \pi\I\, \theta_I Q^I\ ,
\ee
to be extremized over $Q^I$ and $z$. The extremal value of $z$ is proportional to the phase of
the central charge, 
\be
\label{sadz}
z = \I \sqrt{Z/\bar Z}\ .
\ee
Plugging this value back into \eqref{Sigz}, we recover \eqref{FullS}.
The extremization of \eqref{Sigz} with respect to 
$Q^I$ amounts  to a Legendre transform of the Bekenstein-Hawking entropy $S_{BH}(Q)$.
We define the Hesse potential to be the opposite of the Legendre transform of $S_{BH}(Q)$ \cite{LopesCardoso:2006bg},
\be
\label{defHes}
\Sigma(\phi_I) \equiv \langle - S_{BH}(Q)  + \pi \, \phi_I Q^I \rangle_{Q^I}\ ,
\ee
where $Q^I=(q_\Lambda,p^\Lambda)$ includes both the electric and magnetic charges,
and $\phi_I=(\zeta^\Lambda,\tzeta_\Lambda)$ includes both electric and magnetic potentials. 
Like the Bekenstein-Hawking entropy, the 
Hesse potential is homogeneous 
of degree two, and can be evaluated by using
the ``dual attractor equations''\footnote{The terminology is only meant to emphasize the similarity
of \eqref{dualatt} and \eqref{att}, and does not imply any physical attractor behavior for the potentials
$\phi_I$.}
 (\cite{Pioline:2006ni}, Ex. 8)
\be
\label{dualatt}
 \Im(\cF_I)=- \phi_I \quad \Rightarrow \quad \Sigma(\phi) = \frac{\I\pi}{4} \cF_I \bar \cF^I  \ .
\ee
Comparing  \eqref{dualatt} and \eqref{att} we conclude from that the Hesse potential is a positive function, equal to the Bekenstein-Hawking entropy function after replacing $Q^I$ with $\phi_I$. 
Applying \eqref{defHes} to \eqref{Sigz}, we obtain 
\be
\label{SQ}
\langle \Sigma(Q,g_s,t^a,z)  \rangle_{Q} = \frac{1}{g_s^2} 
\Sigma\left( \I \, e^{K/2} (\cF_I\, z^{-1} - \bar \cF_I z) -2 \I g_s \theta_I \right)\ ,
\ee
to be further extremized over $z$. Substituting \eqref{sadz} in \eqref{SQ} and expanding
to leading order in $g_s$, we conclude that $\kappa$ is positive. Thus, the ambiguity 
of the D-instanton series is comparable to the  expected contributions from KK5 or 
NS5-brane instantons wrapped on $X$.

Our second (related) task  is to resum the D-instanton series in the region where 
$\Re(e^K/g_s^2)>0$ 
by generalizing the Borel-Laplace resummation method to the case of asymptotic series
with Gaussian growth.
For this purpose, we represent the exponential of the  Bekenstein-Hawking entropy
as a contour integral  
\be\label{Hesse-pot}
e^{S_{BH}(Q)}  \sim \int \de \phi_I \, e^{-\Sigma(\phi) + \pi\, \phi_I Q^I}\ ,
\ee
where the variables $e^{\phi_I}$ can be thought of as the 
``Borel plane" variables. In writing \eqref{Hesse-pot} we remain imprecise about
the specific choice of integration contour in the Borel plane, as it cannot 
be fixed without additional input. Here we require only that it selects the same 
saddle point as the Legendre transform \eqref{defHes}, and 
neglect  power corrections to the saddle point approximation.

The D-instanton sum \eqref{singleestim} can now be rewritten as 
\be
\begin{split}
\label{resum}
\cA(g_s,\theta) &= \sum_{Q\in L}  \mu(Q) \, 
e^{-\frac{1}{g_s} \Scl(Q,t^a) + 2\pi\I \theta_I Q^I} \\
&=  \int  \de \phi_I\, e^{-\Sigma(\phi)}  \left( \sum_{Q\in L} a(Q)\, 
 e^{-\frac{1}{g_s} \Scl(Q,t^a)+ 2\pi \I (\theta_I - \frac{\I}2 \phi_I ) Q^I} \right)  \ ,
\end{split}
\ee
where, in the second equality, we exchanged the summation over $Q$ with the integral 
over $\phi_I$, in effect implementing a ``Borel-Gauss'' resummation. 
According to our assumptions, the sum in bracket has now finite radius of convergence 
in $e^{\phi_I}$, but may have singularities away from the origin.  Again, since $\Scl(Q,t^a)$
scales linearly in $Q$, we expect poles at $\phi_I^*+2\I \theta_I \sim 1/g_s$,  
leading to ambiguities of order $e^{-\Sigma(\phi^*)}\sim e^{-1/g_s^2(1+\cO(g_s))}$
in the coupling $\cA$.

Just as in \eqref{Sigz}, in the case of 3D backgrounds with 8 supercharges
it is convenient to write  the exponential of the 
classical action as a contour integral over a ``twistor coordinate''~$z$  \cite{Neitzke:2007ke},
\be
\begin{split}
\cA(g_s,\theta) & \sim
 \int  \de \phi_I \, e^{-\Sigma(\phi)} \int \frac{\de z}{z^{1+\delta}} 
 \sum_{Q\in L} a(Q)\,  \\
 &
\qquad \exp\left[-\left( \frac{\I\pi\,e^{K/2}}{g_s\, z} \cF_I   -  
 \frac{\I\pi\, e^{K/2} z}{g_s} \bar \cF_I  -2\pi\I (\theta_I -\frac{\I}{2} \phi_I) \right)  Q^I
\right]
 \ ,
 \end{split}
\ee
so that electromagnetic charges $Q^I$ now appear linearly in the exponent.
The integral over $z$ is of Bessel type, with a saddle point at \eqref{sadz}, 
and reproduces \eqref{resum} up to 
irrelevant power corrections,  irrespective
of the value of $\delta$. We can now perform a Poisson resummation on $Q^I$, 
\be
\cA(g_s,\theta) \sim
 \int  \de \phi^I \, e^{-\Sigma(\phi)}\,  \int \frac{\de z}{z^{1+\delta}}  \left[ \sum_{M_I\in L^*} 
 b\left( \theta_I  -\frac{\I}{2} \phi_I -\frac{e^{K/2}}{2g_s\, z} \cF_I   +  \frac{e^{K/2} z}{2 g_s} \bar \cF_I  
 - M_I \right) 
 \right] \ ,
\ee
where $b(M_I)$ is the Fourier transform of $a(Q^I)$;  given our assumptions on $a(Q)$,
$b(M)$ is peaked around the origin $M_I=0$. For simplicity, we shall approximate 
$b(M)$ by a Dirac delta function, which would be exact if $a(Q)$ was equal to
a constant. Thus, we
obtain
\be
\cA(g_s,\theta) \sim \int \frac{\de z}{z^{1+\delta}}  \sum_{M_I\in L^*}  \, e^{-\Sigma(\phi_I^*)}\, 
 \ee
where
 \be
\phi_I^*
= -2\I ( \theta_I -  M_I) +\frac{\I\, e^{K/2}}{g_s} \left(  \cF_I  \, z^{-1} -     \bar \cF_I  \, z \right)  
\ee
and the sum only should include terms with $\Re[ \Sigma(\phi^*) ]>0$.
To evaluate $\Sigma(\phi^*)$, we may now use \eqref{dualatt}. For example,
setting $\theta_I =M_I$ and $z=\pm 1$, one finds
\be
\Sigma(\phi_I^*) = 
\frac{\I\pi}{4g_s^2} e^{K} \cF_I \bar \cF^I=2\pi R^2 \tau_2^2 V \ ,
\ee
which is precisely the action of a Kaluza-Klein
monopole  wrapped on the Calabi-Yau threefold $X$. Unfortunately, we are not able
to perform the remaining integral over $z$.
Away from  $\theta_I = M_I$, the quantum numbers $M_I$ give corrections 
of order $g_s$ to the KK5-brane action, and should be interpretable as the charges of 
D-instantons bound to the KK5-brane.  Of course, the classical action misses the
minimal coupling to the NS-axion (or NUT potential) $\sigma$, which implies that the 
instanton responsible for the ambiguity in the Borel resummation should have zero
total KK5-brane number, i.e. correspond to  a supersymmetric bound state of a KK5-brane and an 
anti-KK5-brane. The fact that the Hesse potential (and therefore the Bekenstein-Hawking entropy function) controls the classical action of KK5-brane configurations is an interesting outcome of our
analysis. It is perhaps not unexpected, since $\Sigma$ also controls the \kahler
potential on the twistor space of the three-dimensional moduli space \cite{Neitzke:2007ke},
while radially symmetric KK5-branes can be obtained as certain kind of geodesics on 
this space \cite{Gunaydin:2005mx,Gunaydin:2007bg}.

In general, in addition to the power suppressed corrections to $S_{BH}(Q)$,
which are encoded in the Bekenstein-Hawking-Wald entropy, one expects further
exponentially suppressed corrections.  In the specific example of $\cN=4$ string vacua
in four dimensions, where the dyon degeneracies  are known exactly \cite{Dijkgraaf:1996it}, 
these corrections take the form 
\be
\label{subexp}
\Omega(Q) = \sum_{k=1}^{\infty} \, \Omega_k(Q) \, e^{\frac{1}{k} S_{BH}(Q)}\ ,
\ee
where $\Omega_k(Q)$ is an infinite set of power corrections around each 
exponential term \cite{Banerjee:2008ky}. A similar form is also expected for 
$\cN=2$ black holes, based on the Rademacher expansion  \cite{Dijkgraaf:2000fq,deBoer:2006vg} 
of the elliptic genus of the MSW \cite{Maldacena:1997de}
superconformal field theory. The Borel-Gauss resummation
discussed above can be applied to the terms with $k>1$ upon replacing 
$\Sigma\to k \Sigma$ in the previous derivation, and leads to exponentially 
suppressed corrections of order  $e^{-k/g_s^2}$, characteristic of 
bound states of $k$ KK5-branes. 

As a specific example of the general phenomenon discussed above, we now
discuss the instanton corrections to the hypermultiplet moduli space in type II
theories compactified on a Calabi-Yau three-fold $X$. As reviewed e.g. in 
\cite{Alexandrov:2008gh}, the hypermultiplet space in type IIB string theory
receives instanton corrections from Euclidean D(-1), D1, D3 D5-branes wrapping 
complex cycles in $H^{\rm even}(X,\IZ)$ (or more generally elements in the
derived category of $X$, labelled by charges $Q^I$ in the K-theory lattice $L=K(X)$), 
and from NS5-branes wrapping $X$. The D-instanton corrections, 
to linear order around the one-loop corrected moduli 
space metric, are encoded in the ``contact potential" \cite{Alexandrov:2008gh,Alexandrov:2009zh}
(closely related to the \hk potential
on the Swann bundle over $\cM$ \cite{deWit:2001dj,Alexandrov:2008nk})\footnote{To translate into the notations of  \cite{Alexandrov:2008gh}, recall that in this 4D set-up, $1/g_s^2 = 8 V \tau_2^2 
=e^{-K} \tau_2^2$.} :
\be
\begin{split}
e^{\Phi}=& \frac{1}{16 g_s^2} +\frac{\chi_X}{192\pi} \\
& +\frac{1}{16\pi^2 g_s^2} \sum_{Q}  \, n_Q \, \sum_{m> 0}
\frac{|Z(Q)|}{m}\, \cos\left(2\pi m \, \theta_I Q^I \right)
K_1\left(2\pi m\,  | Z(Q) | / g_s \right)\, + \dots
\end{split}
\label{phiinstfull}
\ee
Comparing to \eqref{singleestim} and using 
$K_1(z)\sim \left(1+\cO(1/z)\right)\, e^{-z} \sqrt{\pi/2z}$, we read-off the D-instanton measure 
\be
\label{mu0hy}
\mu( Q, g_s, t^a) = (1+\cO(g_s))\, 
\frac{|Z(Q)|^{1/2} }{64\pi^2 g_s^{3/2}} \sum_{m|Q^I} m^{-2}\, n_{Q/m}\ .
\ee
When $Q^I$ is a primitive vector in the lattice $L$, only $m=1$ contributes to the sum,
and therefore $\mu\sim n_Q$ up to normalization factors. It is worthwhile to note that
the same sum over divisors appears for D(-1)-instantons in 10 dimensions \cite{Green:1997tv}.
The NS5-brane contributions are not well understood at present, although some 
suggestions have been made \cite{Dijkgraaf:2002ac,Kapustin:2004jm,Pioline:2009qt}.

The D-instanton measure $\mu$ may be related to indexed degeneracies 
of four-dimen-sional BPS black holes as explained above \eqref{muom}. Specifically, the hypermultiplet
moduli space is unaffected by reduction to 3 dimensions on a circle $S_1$ of radius $\tilde R$. 
Under T-duality along that circle, it is identified with the vector multiplet moduli space in type IIA string
theory compactified on the same Calabi-Yau three-fold $X$ times the T-dual $S_1$
of radius $R=1/\tilde R$ . The afore mentioned D-instantons are T-dual
to BPS black holes in 4 dimensions, 
obtained by wrapping D0,D2,D4,D6 branes on complex cycles in the homology 
class $Q\in L$ times the circle $S_1(R)$. Thus, the D-instanton measure $\mu_0$
(after dropping the moduli dependent prefactor in \eqref{mu0hy})
should be equal to the indexed degeneracy of a four-dimensional black hole in the same homology class, or in mathematical terms, to the generalized Donaldson-Thomas invariant \cite{MR2302500,ks} 
$n_Q$ \cite{Alexandrov:2008nk} (for $Q\in H^0+H^2$,
they must reduce to the genus 0 Gopakumar-Vafa invariants of $X$ \cite{RoblesLlana:2006is}).

The divergence of the resulting  D-instanton series \eqref{phiinstfull} has often been raised as an objection against the equality of the instanton measure and the  indexed degeneracy of BPS
black holes, and therefore against the usefulness of the hypermultiplet moduli space 
as a book-keeping device for microscopic degeneracies of BPS black holes \cite{Gunaydin:2005mx}.   
As we have argued in this note, this 
objection is not as fatal as it once seemed: in spite of its Gaussian growth, it is perfectly
consistent to treat the D-instanton series as an asymptotic series, with an inherent ambiguity 
of order $e^{-1/g_s^2}$. This ambiguity is precisely of the correct magnitude to be 
cancelled by KK5-brane contributions to the vector multiplet branch, or by
NS5-brane contributions to the hypermultiplet branch. Realizing this scenario 
will require a far-reaching extension of the framework 
of \cite{MR2302500,ks,Gaiotto:2008cd,Alexandrov:2008gh}  into the
NS5/KK5  sector. In particular, one may wonder whether NS5/KK5-brane contributions are 
themselves Borel summable, or whether yet more exotic effects are still looming behind.

\acknowledgments

It is a pleasure to thank S. Alexandrov, M. Berkooz, J. de Boer, 
G. 't Hooft, G. Moore, A. Neitzke, F. Saueressig and A. Strominger for 
discussions. B.P. is grateful to the Spinoza institute for hospitality
during the completion of this work.  
The research of B.P. is supported in part by ANR (CNRS-USAR)
contract no.05-BLAN-0079-01. 


\providecommand{\href}[2]{#2}\begingroup\raggedright\endgroup

\end{document}